\documentclass{mn2e}
\usepackage{epsf}
\usepackage{graphicx}
\usepackage{subfigure}

\newcommand{\vecdot}{{\,\raise 2.0pt\hbox{\bf .}\,}}
\newcommand{\vcross}{\raise.23ex\hbox{$_\wedge$}}
\newcommand{\spose}[1]{\hbox to 0pt{#1\hss}}
\newcommand{\simlt}{\mathrel{\spose{\lower 3pt\hbox{$\mathchar"218$}}
     \raise 2.0pt\hbox{$\mathchar"13C$}}}
\newcommand{\simgt}{\mathrel{\spose{\lower 3pt\hbox{$\mathchar"218$}}
     \raise 2.0pt\hbox{$\mathchar"13E$}}}
\newcommand{\therefore}{\rlap{\lower 0pt\hbox{${.}\mkern7mu{.}$}}
     {\raise 4.0pt\hbox{$\mkern6mu{.}$}}\  }

\title[Warps in protostellar discs]
{The observable effects of tidally induced warps in protostellar discs}

\author[C. J. Nixon and J. E. Pringle]
{C. J. Nixon$^{1,2}$ and
J.E. Pringle$^{1}$\\ $^1$Institute of Astronomy,
University of Cambridge, Madingley Road, Cambridge CB3 0HA\\ $^2$Theoretical Astrophysics Group, University of Leicester,
Leicester, LE1 7RH}

\date{Accepted 2009 December 23.  Received 2009 December 14; in original form 2009 November 2.}

\volume{000}

\pagerange{\pageref{firstpage}--\pageref{lastpage}} \pubyear{2006}

\begin{document}

\label{firstpage}

\maketitle

\begin{abstract}

  We consider the response of a protostellar disc to a tidally induced
  warp and the resultant changes in the spectral energy distribution
  (SED). We argue that for typical protostellar disc parameters the
  warp is communicated through the disc in a wave-like fashion. We
  find that the main effects of the warp tend to be at large radii ($R
  \simgt 30$ AU) and, for sufficiently small viscosity, can be quite
  long-lived. This can result in non-uniform illumination of the disc
  at these radii and can induce significant changes to the SED at
  wavelengths $\lambda \simgt 100 \mu$m.

\end{abstract}

\begin{keywords}
  accretion, accretion discs; stars: formation; stars: pre-main-sequence 
\end{keywords}

\section{Introduction}

Protostars have circumstellar discs from which they accrete
matter. Once the local accretion rate onto the protostar falls below a certain
value, given approximately by
\begin{equation}
\dot{M} = \frac{R_\ast L_\ast}{GM} = 10^{-7} \left(
  \frac{L_\ast}{L_\odot} \right) \left( \frac{R_\ast}{3 R_\odot}
\right) \left( \frac{M}{M_\odot} \right)^{-1} M_\odot {\rm yr}^{-1},
\end{equation}
where $M$ is the stellar mass, $R_\ast$ the stellar radius and
$L_\ast$ the stellar luminosity, then most of the energy radiated from
the disc comes from re-radiated stellar emission, rather than its own
internally generated accretion energy. Such discs are known as passive
discs and most observed protostellar discs fall into this
category. The nature of the spectrum emitted by the disc depends on
how much of the stellar flux can be intercepted and at what radii this
occurs. A flat, infinitesimally thin disc, has at large radii a
temperature distribution of the form $T_{\rm d} \propto R^{-3/4}$
which gives rise to an SED of the form
$\lambda F_\lambda \propto \lambda^{-4/3}$. Almost all protostellar
SEDs are flatter than this, implying that real discs intercept
substantially more of the stellar flux. For this reason, Kenyon \&
Hartmann (1987; see also the review by Dullemond et al., 2007) argued
that protostellar discs are substantially flared, that is, that the
disc scaleheights $H$ are such that $d(H/R)/dR > 0$. Since then more
sophisticated models of the flaring have been introduced (Chiang \&
Goldreich, 1997; D'Alessio et al., 1999; Chiang et al., 2001;
Dullemond \& Dominik, 2004a,b), together with truncated discs which
display a heated inner rim (Natta et al, 2001; Dullemond et al. 2001;
Muzerolle, et al., 2003; Isella \& Natta, 2005). These models are
generally successful in providing explanations of the observed SEDs in
the near and mid-infrared.

These models, however, all assume the disc to be cylindrically
symmetric, and this seems to be a reasonable assumption in the inner
regions (Monnier et al., 2006). Stars typically form within clusters,
and so even single stars are likely to have undergone close encounters
with neighbouring stars at a young age. Very close encounters are
likely to be dissipative and lead to disc truncation and perhaps even
capture and binary star formation (Clarke \& Pringle 1991, 1993; Hall,
Clarke \& Pringle, 1996; Boffin et al., 1998; Watkins et al., 1998a,b;
Pfalzner et al., 2005; Moeckel \& Bally, 2006). More distant
encounters occur more frequently, but can still provide the disc with
significant warping. We note that there is observational evidence for
such disc disturbances (Quillen, 2006; Lin et al., 2006; Cabrit et
al., 2006) and for the warping of the disc at large radius (Hughes et
al., 2009).

Viscosity in the outer regions of protostellar discs is likely to be
small, and typical figures of the dimensionless viscosity parameter
$\alpha$ (Shakura \& Sunyaev, 1973) in the range $10^{-2} - 10^{-4}$
are thought likely, e.g. Terquem, 2008. This is less than the disc
thickness ratio $H/R$, and in this case such warps are propagated as
waves (Papaloizou \& Lin, 1995). With such low viscosities, these
waves can exist for a significant fraction of the protostellar
lifetime (Section~\ref{estimates}).

In this paper, we investigate the effects of these waves in a very
simplified manner. We do not attempt to model perturbations to the
structure or SEDs of discs with flares or other configurations. Rather we
consider the effect on the shape and on the resultant SEDs emitted by a
disc which is initially thin and flat. Although this is not likely to
enable us to provide realistic fits to any observations, it does
enable us to see at what wavelengths any complications deriving from
such perturbations might manifest themselves, and to provide an
estimate of their relative magnitude. We note that the effect of
warping on the SED of a protostellar disc has been considered
previously for the case of a disc in a misaligned binary system by
Terquem \& Bertout (1993, 1996), but that they do not consider the
dynamic response of the disc, nor its subsequent evolution towards
alignment (e.g. Bate et al., 2000).  

In Section~\ref{estimates}, we provide estimates of the periods and
lifetimes of warp-like perturbations in a protostellar disc. In
Section~\ref{warps} we compute the time-dependent behaviour of warps
which can be generated using a simple model of the flyby of a
perturbing star. In Section~\ref{SED} we compute the temperature
distributions and resulting SEDs for these disc models. We discuss
the implications of our results in Section~\ref{discussion}.

\section{Warp propagation and lifetime estimates}
\label{estimates}

We consider a protostellar disc in orbit around a star of mass
$M$. For a Keplerian disc the angular velocity of disc material is
$\Omega = (GM/R^3)^{1/2}$. We take the disc semi-thickness to be $H$,
so that at radius $R$
\begin{equation}
\frac{H}{R} = \frac{c_s}{R \Omega},
\end{equation}
where $c_s(R)$ is the local disc sound speed (Pringle, 1981). Typical
values of $H/R$ are expected to be around 0.05 -- 0.1 (Bell et al.,
1997; Terquem, 2008), independent of radius $R$. Typical estimates for
the dimensionless viscosity parameter $\alpha$ for such discs are in
the range $\alpha \approx 10^{-2} - 10^{-4}$ (Hartmann 2008). Thus for
protostellar discs we expect that $\alpha < H/R$, and therefore that
warps propagate through them in a wave-like manner (Papaloizou \& Lin,
1995). The propagation speed for these waves is (Papaloizou \& Lin,
1995; Pringle, 1999; Lubow, Ogilvie \& Pringle, 2002)
\begin{equation}
v_{\rm w} = \frac{1}{2} c_{\rm s}.
\end{equation}

Suppose that a warp is induced in the outer disc regions by the fly-by
of some other stellar object. If the relative velocity of the
encounter is $V$ and the distance of closest approach is $D$ then the
characteristic frequency of the induced warp-like disturbance is
\begin{equation}
\omega_{\rm w} \approx \frac{V}{D},
\end{equation}
and the wavelength of the disturbance is
\begin{equation}
\lambda_{\rm w}(R)  \approx \frac{ v_{\rm w}}{\omega_{\rm w}} \approx
\frac{c_{\rm s}(R)}{2 V} D.
\end{equation}

We expect the warp to propagate inwards to a radius $R_{\rm crit}$ at
which $R = \lambda(R)$. At that radius the propagating warp wave is
reflected, and within that radius the disc is flat, but tilted by the
warp. This is because the wavelength of the warp is greater than the
size of the disc in the inner regions and therefore cannot communicate
the warp. However angular momentum is still communicated into the
inner disc, so the disc tilts without warping.  If $H/R \approx$
const., independent of radius, then we find that at radius $R_{\rm
  crit}$
\begin{equation}
\Omega(R_{\rm crit}) \approx \frac{2R}{H} \omega_{\rm w}.
\end{equation}
If we make the further assumption that the fly-by is parabolic, so
that to a first approximation 
\begin{equation}
\omega_{\rm w}^2 \approx \frac{GM}{D^3},
\end{equation}
then we find that
\begin{equation}
\frac{R_{\rm crit}}{D} \approx \left( \frac{H}{2R} \right)^{2/3}.
\end{equation}
Thus we expect that while the inner disc regions $R < R_{\rm crit}$
will have a time-varying tilt, it is only the outer disc regions $R >
R_{\rm crit}$ that will show a significant warp. As a numerical
example, if we assume that the fly-by passes the disc at a distance of
$D = 250$ AU, and $H/R = 0.1$ then we have $R_{\rm crit} \approx 34$ AU. 

The time taken for the warp to propagate to the disc centre and back
is
\begin{equation}
t_{\rm w} \approx \frac{4 R_{\rm out}}{c_{\rm s}},
\end{equation}
where $R_{\rm out}$ is the radius of the outer disc edge. Equivalently
this may be written as
\begin{equation}
t_{\rm w} \approx \frac{2}{\pi} . \frac{R}{H}. P_{\rm out},
\end{equation}
where $P_{\rm out}$ is the orbital time at the outer disc edge. Thus
for an outer disc edge at 100 A.U., and 
\begin{equation}
\label{Rcrit}
H/R \approx 0.1, 
\end{equation}
the warp crossing timescale is around $10^4$ yr.

The decay timescale for the warp is given approximately by (Lubow et
al., 2002)
\begin{equation}
t_{\rm damp} \approx \frac{P_{\rm out}}{2 \pi \alpha}.
\end{equation}
Thus for a disc with $R_{\rm out} = 100$ AU, we would expect the warp
to last for a time of around $1.6 \times 10^6 (\alpha/10^{-4})^{-1}$ yr.

\section{Warp generation - numerical results}
\label{warps}

\subsection{The warp equations}

We model the generation and propagation of the warp using the
linearised equations derived by Lubow \& Ogilvie (2000). These
describe the warp in terms of the local unit tilt vector ${\bf l}(R,t)$
which is a function of radius and time. For a Keplerian disc these are
\begin{equation}
\label{LO1}
\Sigma R^2 \Omega \frac{\partial {\bf l}}{\partial t} = \frac{1}{R}
\frac{\partial {\bf G}}{\partial R} + {\bf T},
\end{equation}
and
\begin{equation}
\label{LO2}
\frac{\partial {\bf G}}{\partial t} + \alpha \Omega {\bf G} = \frac{P
  R^3 \Omega}{4} \frac{\partial {\bf l}}{\partial R}.
\end{equation}
Here $\bf G$ is the internal torque which acts to realign tilted disc
annuli. $\bf T$ is the external torque acting on the disc which in
this case is caused by the stellar fly-by. In addition $\Sigma$ is the
disc surface density, and $P = \int p \, dz = \Omega^2 \Sigma H^2$ is
the vertically integrated pressure. We note that these equations do
not make allowance for dispersive or non-linear effects in the wave
propagation (Ogilvie 2006). For the waves we discuss here dispersive
effects may be of marginal significance, but only at the outer disc
edge. However for the large amplitudes required to substantially change
the SED, non-linear effects may well play a role. However, to take
these effects properly into account would require a full
hydrodynamical model of the disc, which is beyond the scope of the
current investigation.

To solve these equations numerically we adopt the approach of Lubow et
al. (2002). The disc is taken to lie in the $XY$--plane, so that for
small tilts we have to first order
\begin{equation}
{\bf l} = (l_x, l_y, 1),
\end{equation}
where $l_x, l_y \ll 1$. The torques $\bf G$ and $\bf T$ lie in the
$XY$--plane.  We can then use $W = l_x + i l_y$, $G = G_x + i G_y$ and
$T = T_x + i T_y$ as complex representations of $\bf l$, $\bf G$ and
$\bf T$, respectively. We then replace the quantities $W$ and $G$ by
the quantities $A(R,t)$ and $D(R,t)$ defined by
\begin{equation}
W = - \frac{D^\ast}{R \Omega^2},
\end{equation}
and
\begin{equation}
G = \frac{1}{2} \Sigma H^2 R^2 \Omega A^\ast,
\end{equation}
where the asterisk denotes a complex conjugate.

Equations~\ref{LO1} and~\ref{LO2} then become
\begin{equation}
\label{EV1}
\frac{\partial D}{\partial t} = - \frac{c_{\rm s}^2}{2} \left[
  \frac{1}{\Sigma R^{1/2} H^2} \frac{\partial}{\partial R} ( \Sigma
  R^{1/2} H^2 A) \right] - \frac{\Omega}{R \Sigma} T,
\end{equation}
and
\begin{equation}
\label{EV2}
\frac{\partial A}{\partial t} = - \alpha \Omega A - \frac{1}{2}
\frac{ \partial D}{\partial R} - \frac{D}{R}.
\end{equation}

We truncate the disc at $R=R_{out}$
and do not allow for dissipation at the outer boundary. This is because the wavelength of the warp is long in comparison to any
smoothing of the outer edge and so the warp will only see it as a sharp
edge and reflect back inwards.

\subsection{External torque due to fly-by}

We now obtain an expression for the torque density ${\bf T}(R,t)$
exerted upon an annulus of the disc during a stellar flyby. The torque
density exerted on the annulus at radius R with tilt vector $\bf l$ by
a mass of $M_2$ at position vector ${\bf R}_{\rm b}$ is given to first
order in $R/R_{\rm b}$ by (Lubow \& Ogilvie, 2000)
\begin{equation}
{\bf T} = \frac{GM_2}{2 R_{\rm b}^4} \left[ b_{3/2}^{(1)} \left(
    \frac{R}{R_{\rm b}} \right) \right] \Sigma \, R ({\bf R}_{\rm b}
  \vecdot {\bf l})({\bf R}_{\rm b} \vcross {\bf l}),
\end{equation}
where $b_{3/2}^{(1)}$ is the Laplace coefficient. We approximate this
in the form
\begin{equation}
{\bf T} = \frac{3GM_2\Sigma R^2}{(R_{\rm b}^2 + R^2)^{5/2}}  ({\bf R}_{\rm b}
  \vecdot {\bf l})({\bf R}_{\rm b} \vcross {\bf l}).
\end{equation}
We note that since we are using the approximation of only considering
linear warp waves, it is only the $T_x$ and $T_y$ components which are
relevant for the computations here.

\subsection{ The disc shape and its evolution}

We start with a disc which has an outer radius of $R_{\rm out} = 100$
AU. We take the inner radius to be at $R_{\rm in } = 0.01 $AU, and
note that this is much less than the expected value of $R_{\rm crit}$
(equation~\ref{Rcrit}) within which we expect the disc to be tilted
but not warped.

We take the surface density of the disc to be a power-law of the form
\begin{equation}
\Sigma(R) \propto R^{-1},
\end{equation}
(c.f. Andrews et al., 2009; Isella, Carpenter \& Sargent, 2009) and
take $H/R = 0.1 =$ const. This then implies that the sound speed is a
power-law of the form:
\begin{equation}
c_{\rm s} \propto R^{-1/2}.
\end{equation}
We apply a small amount of damping by taking $\alpha = 10^{-4}$.  We
note that for these disc properties, we expect the warp amplitude to
vary in such a manner as to keep the flux ($F_\perp$) of angular momentum
associated with the warp constant, where
\begin{equation}
F_\perp \propto |W| (GMR)^{1/2} \Sigma R.
\end{equation}
For the disc properties we have chosen this corresponds to $|W|
\approx$ const., so that the vertical local disc displacement due to
the warp is roughly proportional to radius.

The disc evolution equations~\ref{EV1} and~\ref{EV2} are evolved
numerically using a leapfrog scheme. Boundary conditions are chosen to
ensure that the inner and outer edges of the disc are stress-free.

To illustrate the possibilities we consider two idealised flybys. We
take the orbit of the perturber to lie along a straight line, constant
velocity path. In practice, of course, the interaction is more
dynamical than this (e.g. Cabrit et al., 2006; Moekel
\& Bally, 2006), however we can, to first order, obtain
the correct warping structure for the disc. Thus in both cases we take
the path to be of the form
\begin{equation}
{\bf R}_{\rm b} = {\bf a}  + {\bf V} (t-t_0).
\end{equation}
Here $t_0$ is some arbitrary time at which the perturber is at
position vector $\bf a$.

\subsubsection{Model 1}

For Model 1 we take the flyby to lie along a straight
line perpendicular to the disc of the form
\begin{equation}
{\bf R}_{\rm b} = (300, 0, V(t-t_0)) {\rm AU}.
\end{equation}
Here $t_0$ is the time of closest approach when the perturber is a
distance of $a = |{\bf a}| = 300$ AU from the disc centre, and 200 AU
from the disc edge. As can be seen the velocity $\bf V$ is in the
$Z$--direction. We take its magnitude $V$ to correspond to the escape
velocity from the point of nearest approach, so that
\begin{equation}
  V = \left( \frac{2GM}{a} \right)^{1/2} = 2.44 \left( \frac{M}{M_\odot}
  \right)^{1/2} \left( \frac{a}{300 {\rm AU}} \right)^{-1/2} {\rm km \, 
    s}^{-1}.
\end{equation}

Because the path of the perturber lies solely in the $XZ$--plane, the
disc annuli are tilted only about the $Y$--axis. Thus we can envisage
the disc warp by considering a cut through the disc in the
$XZ$--plane.

The results for Model 1a for which we take the mass of the perturber
to be $M_2 = 1 M_\odot$ are shown in Figure~\ref{fig:2D for model
  1a}. At time $t=0$ the disc is assumed unperturbed and the perturber
has a $z$--coordinate of $z = -1000$ AU, and thus $t_0 \approx 2000$
yr. The outer edge of the disc is initially pulled downwards. Then as
the perturber passes the disc (passing through the initial disc plane, at
time $t=t_0 \approx 2000$ yr) the outer edge of the disc starts to be
pulled upwards. Thus the disc is set oscillating with an amplitude at
the outer edge of around 5 AU. The period of the oscillation is
\begin{equation}
  P = 2 \int^{R_{\rm out}}_0 \frac{dR}{c_{\rm s}(R)} = \frac{4}{3}
  \frac{R_{\rm out}}{c_{\rm s}(R_{\rm out})}.
\end{equation}
For the disc considered here this gives $P \approx 2100$yrs.

\begin{figure}
  \begin{center}
    \subfigure{\label{fig:1a2D-b}
       \includegraphics[width=2in, height=1.4in]{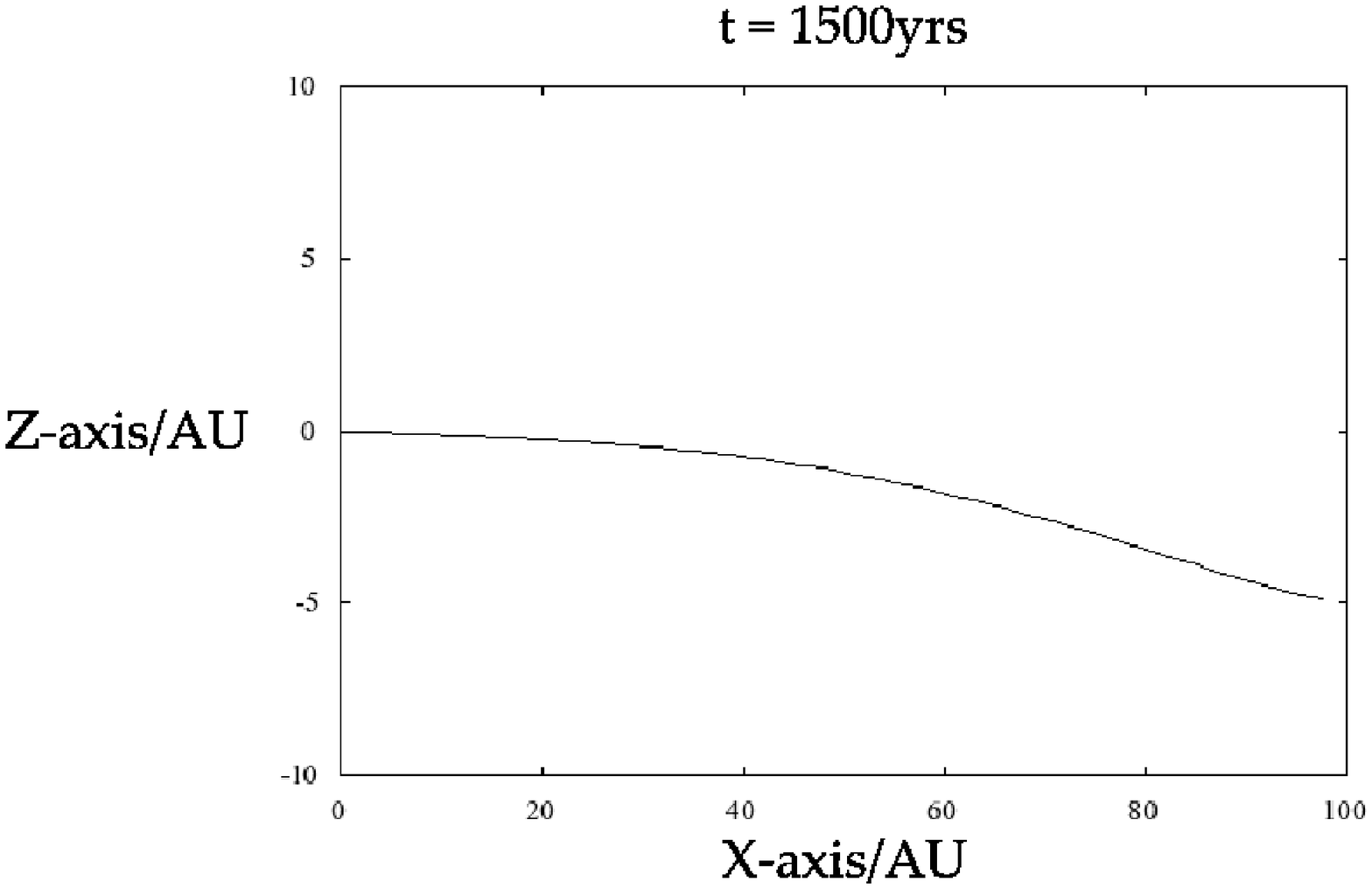}} 
    \subfigure{\label{fig:1a2D-c}
       \includegraphics[width=2in, height=1.4in]{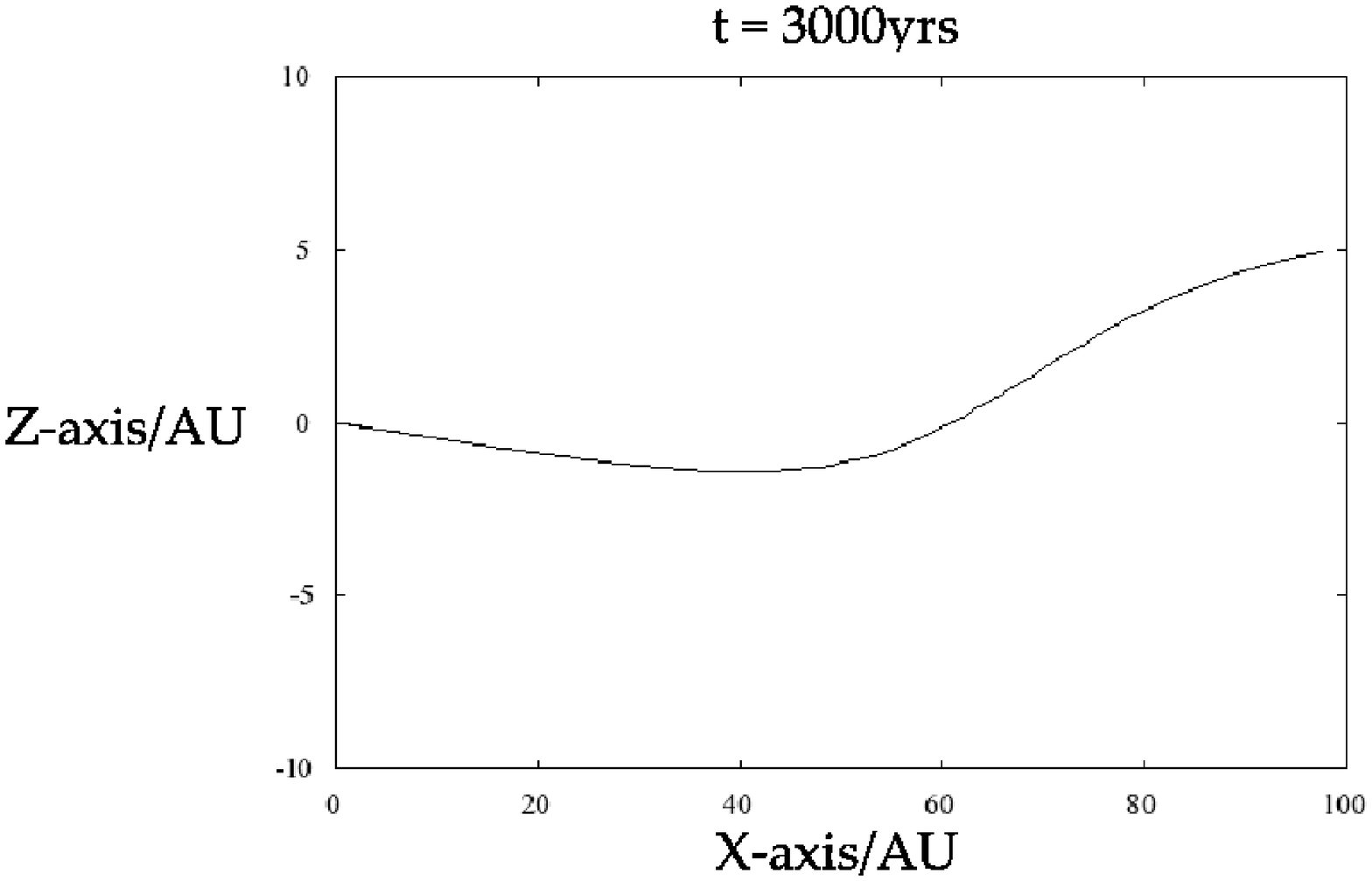}}
    \subfigure{\label{fig:1a2D-d}
       \includegraphics[width=2in, height=1.4in]{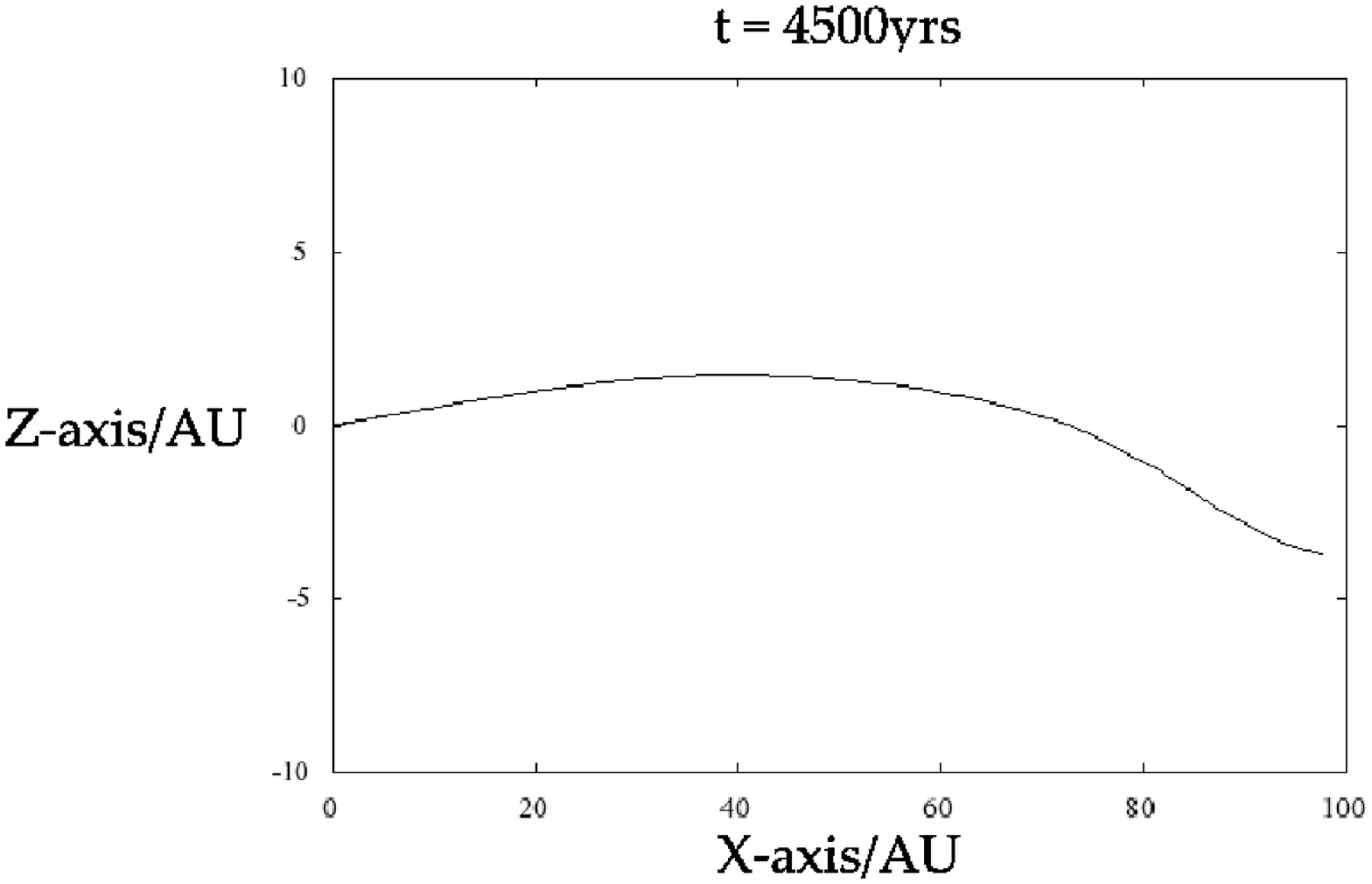}}               
    \subfigure{\label{fig:1a2D-e}  
       \includegraphics[width=2in, height=1.4in]{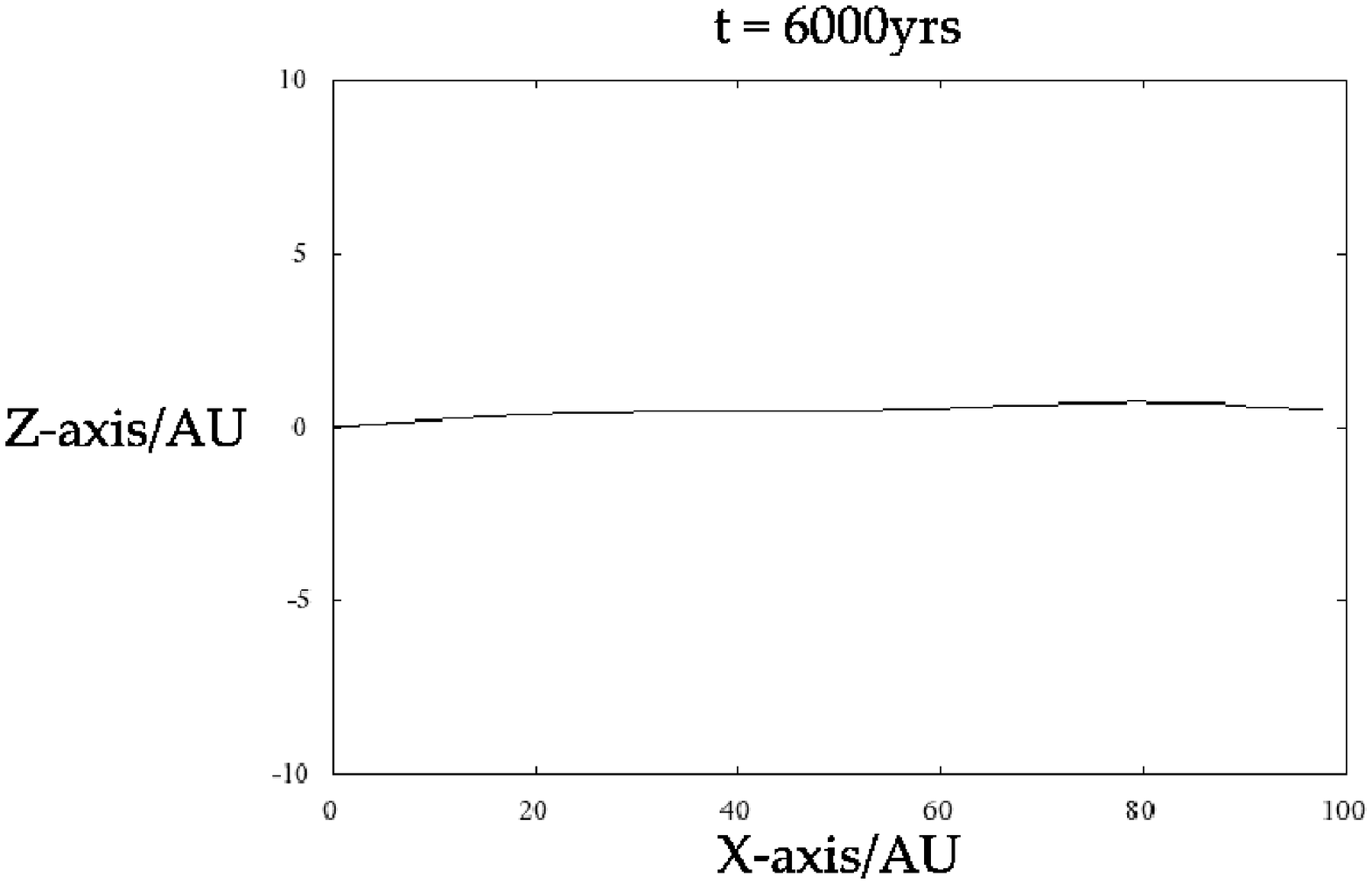}} 
    \subfigure{\label{fig:1a2D-f}
       \includegraphics[width=2in, height=1.4in]{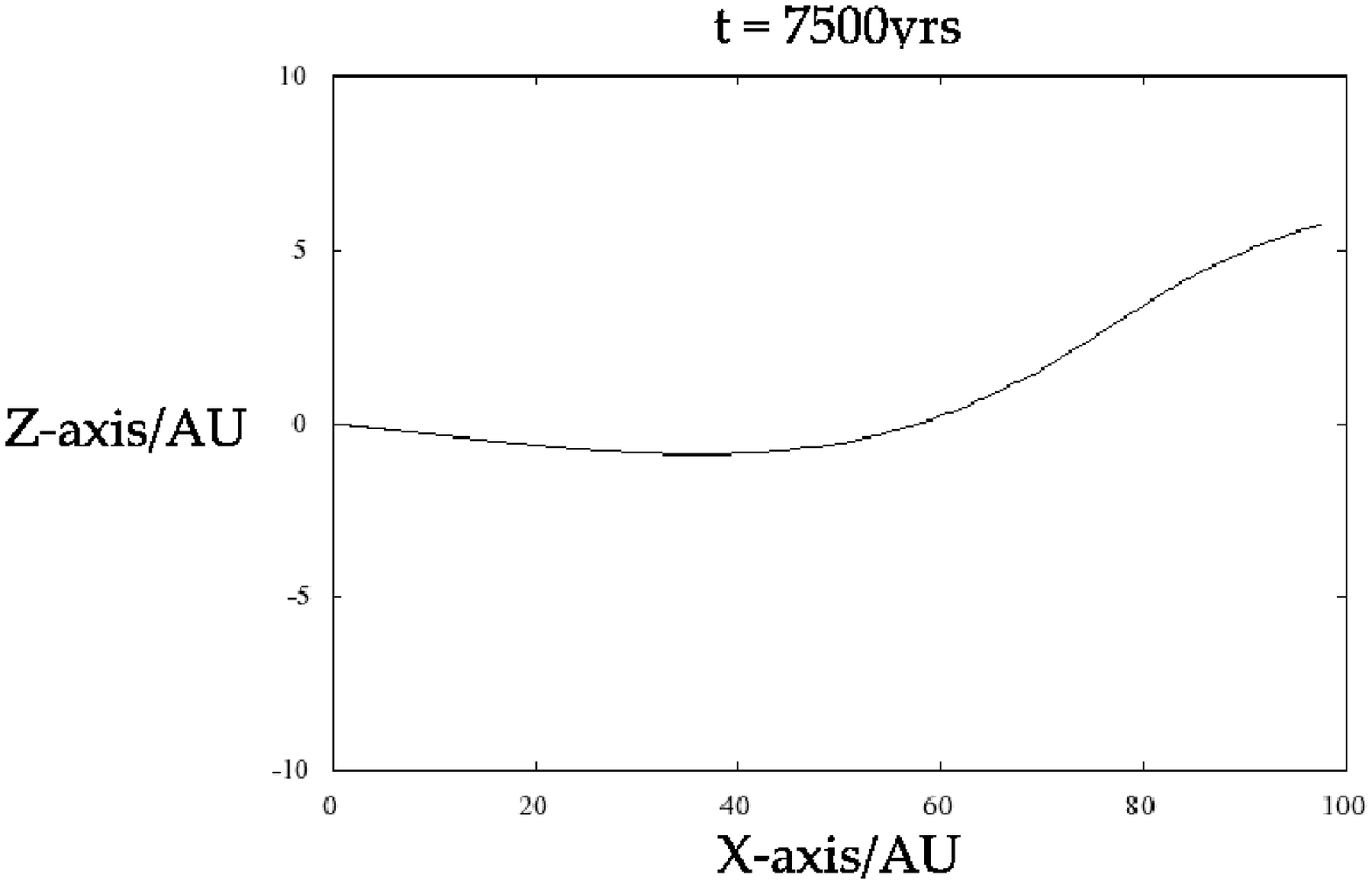}}  
  \end{center}
  \caption{A sample of the $XZ$--profiles of the disc at various times
    for Model 1a, showing the induced warp and its propagation through
    the disc. For time $t=0$, the disc is entirely flat in the $XY$--plane. Here we only show the positive $X$--axis. In the
    $XZ$--plane, the disc is anti-symmetric around $x=0$. The axes are
    in AU. Note that the inner region of the disc ($R \simlt 30$ AU)
    is tilted but not warped.}
  \label{fig:2D for model 1a}
\end{figure}

\begin{figure}
  \begin{center}
       \includegraphics[width=3in, height=2.10in]{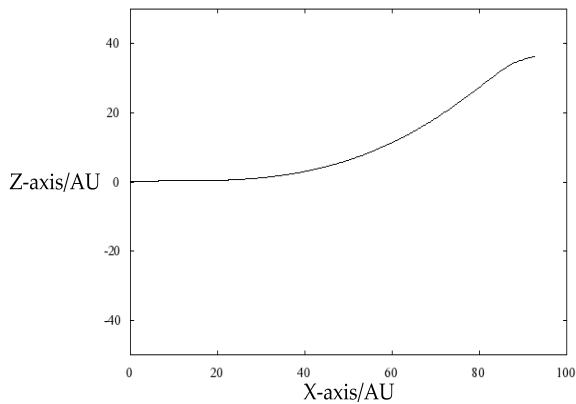}
  \end{center}
  \caption{An example of the $XZ$--profile of the disc for Model 1b,
    showing that in this case the warp is more substantial at the
    outer edge of the disc. The axes are in AU.}
  \label{fig:2D for model 1b}
\end{figure}

For Model 1b we explore a more substantial warping of the disc. For
this we use $a = |{\bf a}| = 200$ AU and we take the mass of the
perturber $M_2 = 10 M_\odot$. We note that this changes the velocity
of the flyby to $V = 9.5$ km s$^{-1}$ . As can be seen in
Figure~\ref{fig:2D for model 1b} this produces an amplitude for the
warp wave of approximately 35 AU.

\subsubsection{Model 2}

In Model 2 we take the path to lie in the direction ${\hat{\bf V}} =
(1,1,1)$, that is, at an angle to the disc plane. The perturber is
initially at position
\begin{equation}
{\bf R}_{\rm b}(t=0) = (-700, -1000, -1000) {\rm AU}.
\end{equation}
The perturber is then given a velocity of $V = |{\bf V}| = 2.7$ km
s$^{-1}$ which corresponds to the escape velocity at the point of
closest approach ${\bf R}_{\rm b} = (200,-100,-100)$ AU. Note that
this path passes through the $XY$--plane at the same point as the path
in Model 1a. Because this path is not in the $XZ$--plane, the torque
acting on disc annuli is now no longer simply in the
$Y$--direction. Thus the disc is not only tilted but also acquires a
twist. For comparison with Model 1a we take $M_2 = 1 M_\odot$. 

\begin{figure}
  \begin{center}
       \includegraphics[width=3.2in, height=2in]{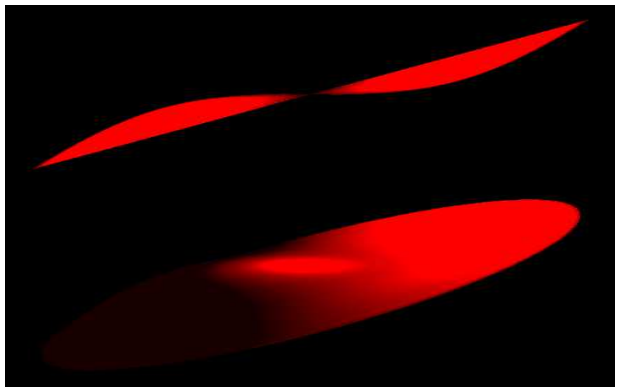}
  \end{center}
  \caption{We show here a representative disc shape in three
    dimensions for Model 1b. The image was generated using the graphics package POVRAY. The top image is the $XZ$ profile of the $100$ AU disc, viewed and illuminated from $(0,-200,0)$. The second image is the same disc but viewed from $(0,-200,50)$, and illuminated from $(0,5,0)$. Because the path of the perturber lies in the $XZ$--plane, each disc annulus is tilted about the
    $Y$--axis. Thus the disc is warped but not twisted.}
  \label{fig:3D for model 1b}
\end{figure}

\begin{figure}
  \begin{center}
       \includegraphics[width=3.2in, height=1in]{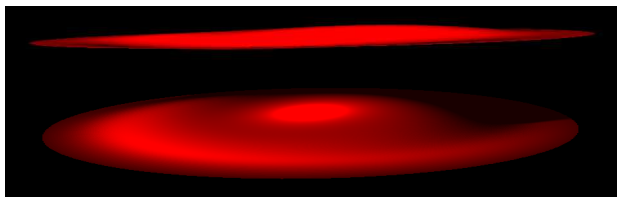}
  \end{center}
  \caption{We show here a representative disc shape in three
    dimensions for Model 2. The image was generated using the graphics package POVRAY. The top image is the $XZ$ profile of the $100$ AU disc, viewed and illuminated from $(0,-200,0)$. The second image is the same disc but viewed from $(0,-200,50)$, and illuminated from $(0,5,0)$. Because the path of the perturber is not perpendicular to the unperturbed disc, the individual disc annuli are tilted about different axes in the $XY$--plane. This means
    that the shape of the disc is twisted about the $Z$--axis. The central region is tilted in one plane and the outer rings are slowly twisted away from this plane, producing the shadowing effects.}
  \label{fig:3D for model 2}
\end{figure}

As the disc is twisted we can no longer view it via a cut through the
$XZ$--plane. To illustrate this, in Figure~\ref{fig:3D for model 1b} we
show the disc in three dimensions, for Model 1b, and compare it to
Figure~\ref{fig:3D for model 2}, which is a typical shape of the disc for
Model 2. The disc shapes for Model 1a and 1b are the same but with differing amplitudes, so we use Model 1b here to illustrate the disc shape more easily.

\section{Disc heating and resultant spectral energy distributions}
\label{SED}

\subsection{Temperature profile}

In order to keep things simple we assume that the disc is thin, and
therefore ignore any flaring. In this way we ensure that any
re-emission of radiation from the disc at large radii (long
wavelengths) is caused predominantly by the warp. As we have noted,
the inner disc (well within the radius $R_{\rm crit} \approx 30$ AU)
remains flat, but with variable tilt. Thus for the inner disc we may
take its surface effective temperature to be $T_{\rm d}(R)$ where

\begin{equation}
T_{\rm d}^4(R) = T_\ast^4 \, \frac{1}{2} \left[ \sin^{-1} \left(
    \frac{R}{R_\ast} \right) - \frac{R_\ast}{R} \sqrt{ 1 - \left(
      \frac{R_\ast}{R} \right)^2 } \right],
\end{equation}
where $T_\ast$ is the stellar effective temperature and $R_\ast$ the
stellar radius (Friedjung, 1985; Kenyon \& Hartmann, 1987).

For the warped part of the disc, we write the local unit vector
representing the disc tilt as
\begin{equation}
{\bf l}(R) = (\cos \gamma \sin \beta, \sin \gamma \sin \beta, \cos
\beta),
\end{equation}
so that the angle $\beta(R)$ represents the tilt of the disc normal,
relative to the $Z$--axis, and the angle $\gamma(R)$ represents the
azimuth of the tilt. Since the warp starts only at a distance $R \geq
R_{\rm crit} \gg R_\ast$ we are able to treat the illumination of the
warped part of the disc by the central star as coming from a point
source. In this case (cf. Pringle, 1996) the disc surface temperature
can be written as
\begin{equation}
  T_{\rm d}^4 = T_\ast^4 \frac{ \left| \sin \phi R \, \frac{\partial
        \beta}{\partial R} - \cos \phi \sin \beta \, R \frac{\partial
        \gamma}{\partial R} \right| } { \left[ 1 + \left( \cos \phi \sin
        \beta \, R \frac{\partial \gamma}{\partial R} - \sin \phi \, R
        \frac{\partial \beta}{\partial R} \right)^2 \right]^{1/2}}.
\end{equation}

This is valid only if self-shadowing of the disc can be ignored. In
practice this needs to be taken into account. Thus this temperature is
applied only to those parts of the disc which are tilted towards the
central star and which have a view of the central star which is
unobscured by other (warped) parts of the disc. Parts of the disc
which are obscured are assumed to have negligible temperature. This
computation is carried out in a straightforward manner as in Pringle
(1997).

Once the temperature distribution of the disc is found, we compute the
emitted disc spectrum simply by assuming that each disc element emits
locally as a black body. For the SED we assume an observer placed on
the Z-axis for simplicity. If the disc were to be viewed from an angle
there would be an apparent reduction in the magnitude of the flux
emitted by the disc. If that angle is increased enough, then the
central star and large portions of the disc would be obscured from the
observer by the warp.

\subsubsection{Model 1}

\begin{figure}
  \begin{center}
       \includegraphics[width=3in, height=1.59in]{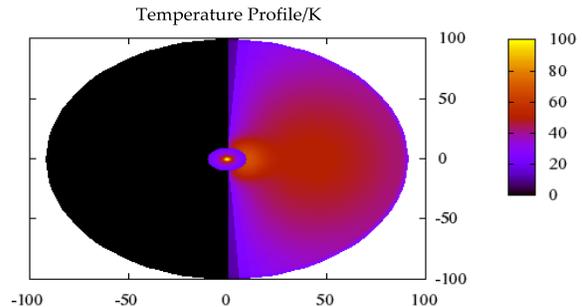}
  \end{center}
  \caption{Temperature profile for the disc in Figure~\ref{fig:3D for
      model 1b}, from Model 1b. The axes are in AU, with the
    temperature coloured by the scale shown in Kelvin. The central
    star (with $T_\ast = 5000$K) and innermost parts of the disc are
    omitted to give a higher resolution over the disc. Note that only
    one half of the disc appears illuminated to an external observer.}
  \label{fig:1bTemp}
\end{figure}

\begin{figure}
  \begin{center}
       \includegraphics[width=3in, height=2.09in]{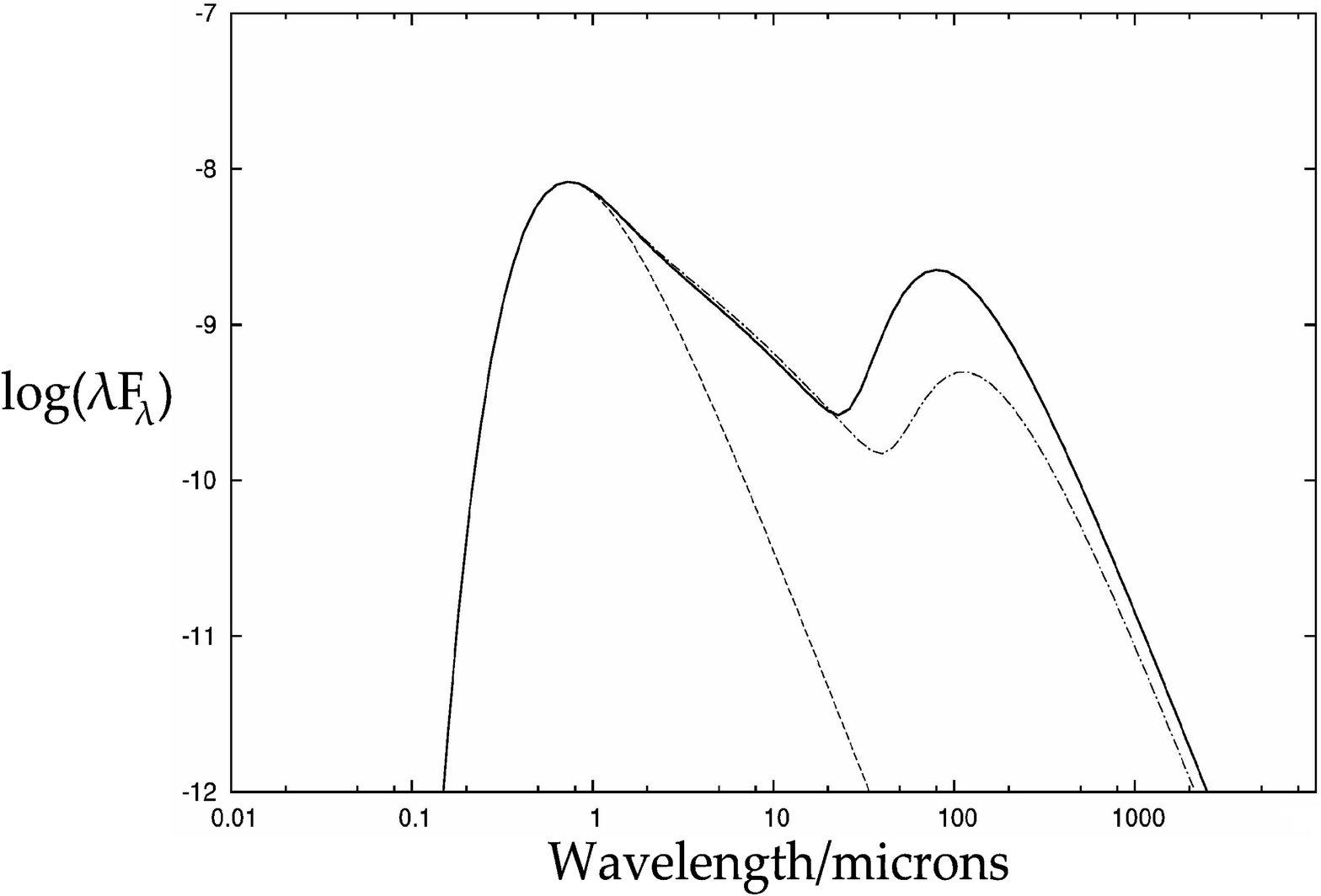}
  \end{center}
  \caption{SED in terms of $\lambda F_\lambda$. The solid line is for the disc in
    Figure~\ref{fig:3D for model 1b}, from Model 1b. The dot-dashed line is a typical SED from       Model 1a, and the dashed line represents the star which is taken to be a black body with
    $T_\ast = 5000$ K. The contribution from the star peaks at about 0.6 $\mu$m.  The inner disc
    has a similar maximum temperature and a log wavelength dependence
    of the form $\lambda F_\lambda \propto \lambda^{-4/3}$. The
    contribution from the upturned part of the outer disc peaks at
    around 100 $\mu$m. As the warp is more substantial in Model 1b, the flux in this peak is        greater than the flux in the corresponding peak for Model 1a. As is to be expected, both        disc peaks have less magnitude than the primary peak, with Model 1b having a disc peak flux that is around 5 times less than the primary peak flux.}
  \label{fig:1bSED}
\end{figure}

In Figure~\ref{fig:1bTemp} we show a representative temperature
distribution found for Model 1b, and in Figure~\ref{fig:1bSED} the
resultant spectral energy distribution (as viewed by a distant
observer in the direction of the $Z$--axis) from the disc structure
shown in Figure~\ref{fig:3D for model 1b}. For comparison we also plot, on figure~\ref{fig:1bSED}, an SED from Model 1a and the contribution to the SED from the central star. Because of the induced warp, the SED displays a
secondary peak at around 100 $\mu$m. For shorter wavelengths the
radiation comes from the stellar surface and the flat illuminated part
of the inner disc. The
magnitude of the warp is larger in Model 1b, creating a more
substantial flux from the warped disc at around 100 $\mu$m as opposed to the weaker flux from Model 1a. In Model 1 all disc
annuli are tilted about the same axis, therefore outside the central region
(taken to be at the reflection radius), only one half of the disc
appears illuminated.

\begin{figure}
  \begin{center}
    \subfigure{\label{fig:shadow1}
       \includegraphics[width=3in, height=2.25in]{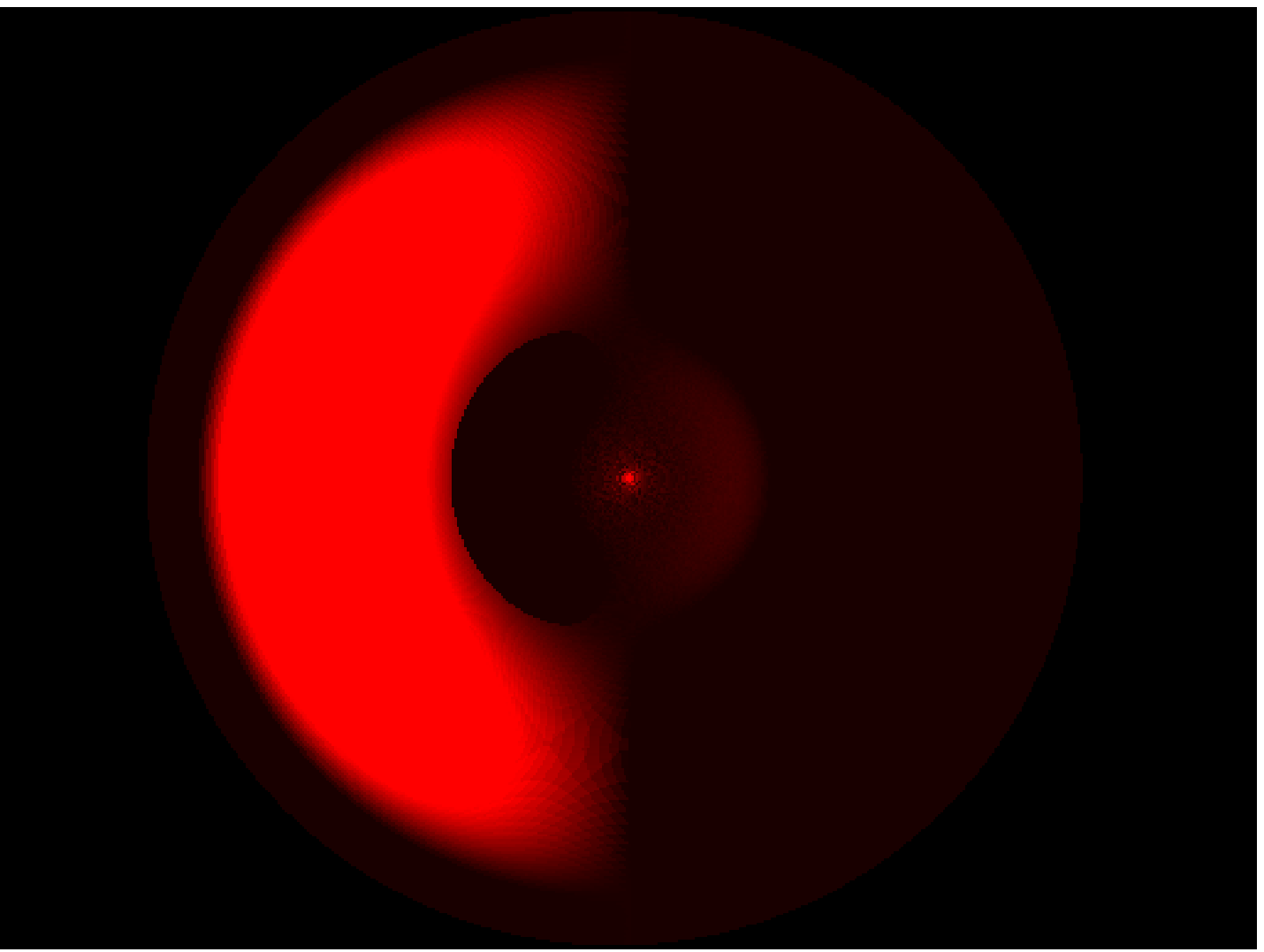}}
    \subfigure{\label{fig:shadow2}
       \includegraphics[width=3in, height=2.25in]{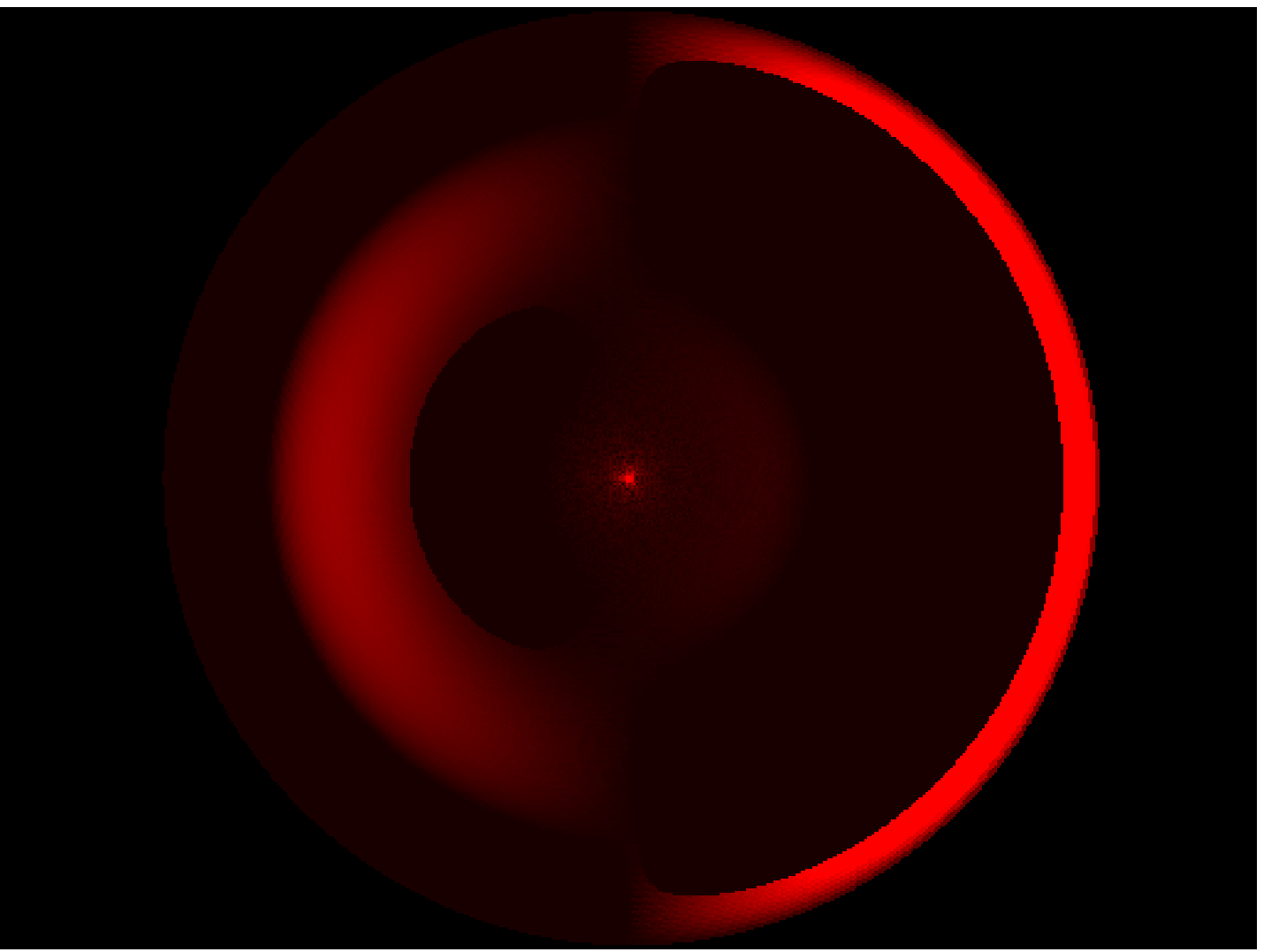}} 
    \subfigure{\label{fig:shadow3}
       \includegraphics[width=3in, height=2.25in]{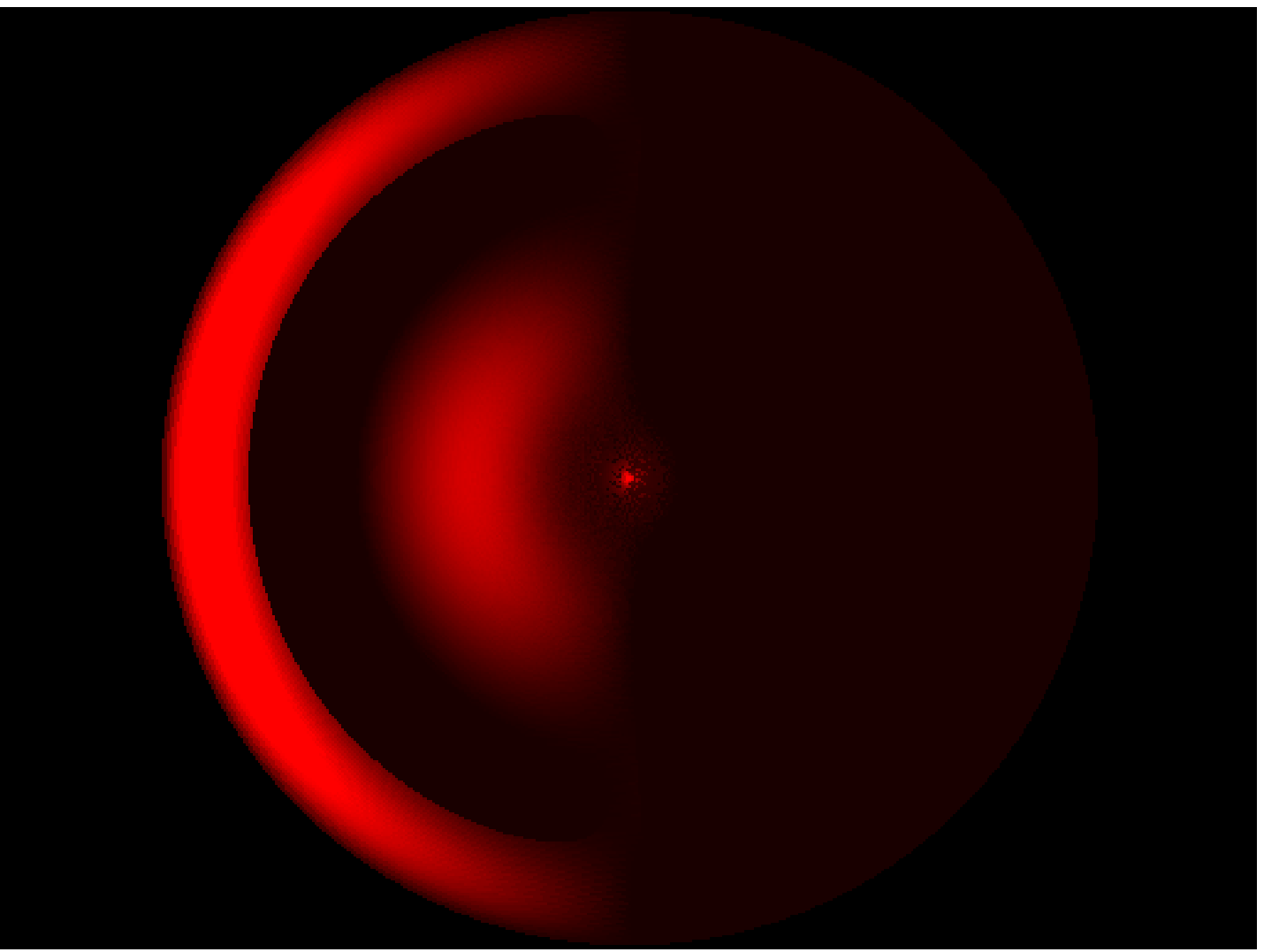}}
  \end{center}
  \caption{Some examples of the complicated illumination profiles that
    arise due to the self-shadowing as the warp propagates through the
    disc. Note that the characteristic property of
    an illuminated warp is that the disc surface brightness is
    lop-sided with respect to the central star. The images were generated by the graphics package POVRAY, with the source of illumination on the central stars surface. The discs here have a radius of $100$ AU.}
  \label{fig:shadow}
\end{figure}

In general the disc illumination can be more complicated than this
because of self-shadowing, and we illustrate some possibilities in
Figure~\ref{fig:shadow}. 

\subsubsection{Model 2}

\begin{figure}
  \begin{center}
       \includegraphics[width=3in, height=1.59in]{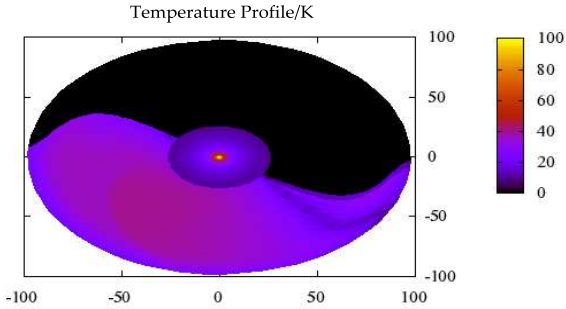}
  \end{center}
  \caption{Temperature profile for the disc in Figure~\ref{fig:3D for
      model 2}, from Model 2. The axes are in AU, with the colour
    scale giving the temperature in Kelvin. The central star (with
    $T_\ast = 5000$K) and innermost parts of the disc are omitted to
    give a higher resolution over the disc. Note that when the path of
    the perturber is not perpendicular to the unperturbed disc plane,
    the disc is twisted and so the pattern of illumination displays an
    apparent spiral pattern.}
  \label{fig:2Temp}
\end{figure}

\begin{figure}
  \begin{center}
       \includegraphics[width=3in, height=2.11in]{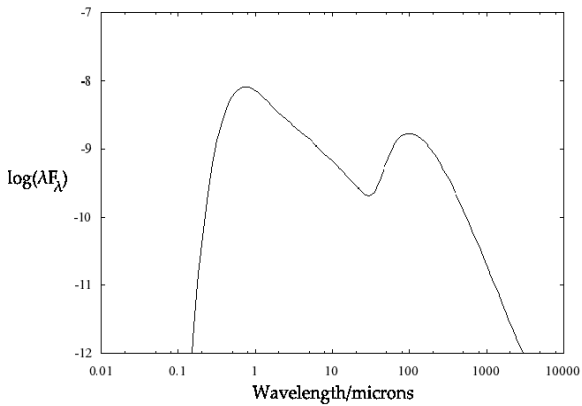}
  \end{center}
  \caption{SED in terms of $\lambda F_\lambda$ for the disc in
    Figure~\ref{fig:3D for model 2}, from Model 2.}
  \label{fig:2SED}
\end{figure}

As the flyby in Model 2 is misaligned with the original plane of the
disc, the resultant disc is both tilted and twisted. This can be seen
in Figure~\ref{fig:2Temp} which shows a typical temperature profile
for Model 2. This exhibits a spiral pattern in the illumination profile. The resultant SED, Figure~\ref{fig:2SED}, is not changed
greatly by this, which is to be expected as the disc intercepts a
similar amount of stellar flux when both tilted and twisted as opposed
to just being tilted. The contribution of the heated outer disc again
peaks at around 100 $\mu$m.

\section{Discussion}
\label{discussion}

We have considered the response of a protostellar disc to a tidally
induced warp, and the resultant change in the spectral energy
distribution (SED). We have used a simplified analysis in order to
emphasise the relevant implications.

For the low viscosities currently envisaged for the outer disc regions
we note that a tidally induced warp is propagated as a wave, and that
such a wave can be comparatively long lasting (in our case more than a
Myr for $\alpha = 10^{-4}$). The warp is largest in the outer regions,
and within some critical radius (in our case $R_{\rm crit} \sim 30$
AU) the disc remains flat but has a tilt which is variable on a
timescale of order the propagation timescale of the warp (here a few
thousand years). Note that any collimated outflow, or jet, originating
from the inner disc regions might be expected therefore to vary
direction on this timescale.

In reality the passive disc that we are considering here will be
flared. Flaring of the disc contributes to the SED at wavelengths
longer than a few $\mu$m. In this paper, in order to emphasise the
effects of the warp, we have assumed the disc to be locally thin when
computing the SED. In practice the two effects, that of the flare and
that of the warp, must be combined.

From our computations, in agreement with Terquem \& Bertout (1993,
1996), we conclude that the warp is only likely to affect the disc SED
at wavelengths of around 100 $\mu$m and longer, but that the effects
can be substantial. We also note that warp-induced variations in disc
illumination have the effect of making the disc appear asymmetric with
respect to the central star at radii $R \simgt R_{\rm crit}$. Further,
a twist in the disc, induced by a skewed flyby, can have the effect of
introducing a spiral component into the disc illumination.

\section{Acknowledgments} 

We thank Cathie Clarke for useful discussions on the topic. We also thank the referee for useful feedback.

\label{lastpage}


\begin{thebibliography}{}

\bibitem{} Andrews, S.M., Wilner, D.J., Hughes, A.M., Qi, C.,
  Dullemond, C.P., 2009, ApJ, 700, 1502

\bibitem{} Bate, M.R., Bonnell, I.A., Clarke, C.J., Lubow, S.H.,
  Ogilvie, G.I., Pringle, J.E., Tout, C.A., 2000, MNRAS, 317, 773

\bibitem{} Bell, K.R., Cassen, P.M., Klahr, H.H., Henning, T., 1997,
  ApJ, 486, 372

\bibitem{} Boffin, H.M.J., Watkins, S.J., Bhattal, A.S., Francis, N.,
  Whitworth, A.P., 1998, MNRAS, 300, 1189

\bibitem{} Cabrit, S., Pety, J., Pesenti, N., Dougados, C., 2006,
  A\&A, 452, 897

\bibitem{} Chiang, E.I., Goldreich, P., 1997, ApJ, 490, 368

\bibitem{} Chiang, E.I., Joung, M.K., Creech-Eakman, M.J., Qi, C.,
  Kessler, J.E., Blake, G.A., van Dishoeck, E.F., 2001, ApJ, 547, 1077 

\bibitem{} Clarke, C.J., Pringle, J.E., 1991, MNRAS, 249, 584

\bibitem{} Clarke, C.J., Pringle, J.E., 1993, MNRAS, 261, 190

\bibitem{} D'Alessio, P., Calvet, N., Hartmann, L., Lizano, S.,
  Cant\'o, J., 1999, ApJ, 527, 893

\bibitem{} Dullemond, C.P., Dominik, C., Natta, A., 2001, ApJ, 560,
  957

\bibitem{} Dullemond, C.P., Dominik, C., 2004a, A\&A, 417, 159

\bibitem{} Dullemond, C.P., Dominik, C., 2004b, A\&A, 421, 1075

\bibitem{} Dullemond, C.P., Hollenbach. D., Kamp, I., D'Alessio, P.,
  2007, in Protostars \& Planets V, ed. B. Reipurth, D. Jewitt \&
  K. Keil (Tucson, AZ: Univ. Arizona Press), 555

\bibitem{} Friedjung, M., 1985, A\&A, 146, 366

\bibitem{} Hall, S.M. Clarke, C.J., Pringle, J.E., 1996, MNRAS, 278,
  303
  
\bibitem{} Hartmann, L., 2008, Physica Scripta, 130, 4012 

\bibitem{} Hughes, A.M., Andrews, S.M., Espaillat, C., Wilner, D.J.,
  Calvet, N., D'Alessio, P., Qi, C., Williams, J.P., Hogerheijde,
  M.R., 2009, ApJ, 698, 131

\bibitem{} Isella, A., Carpenter, J.M., Sargent, A.I., 2009, ApJ, 701, 260 

\bibitem{} Isella, A., Natta, A., 2005, A\&A, 438, 899

\bibitem{} Kenyon, S.J., Hartmann, L., 1987, ApJ 323, 714

\bibitem{} Lin, S.-Y., Ohashi, N., Lim, J., Ho, P.T.P., Fukagawa, M.,
  Tamura, M., 2006, ApJ, 645, 1297

\bibitem{} Lubow, S.H., Ogilvie, G.I., 2000, MNRAS, 538, 326

\bibitem{} Lubow, S.H., Ogilvie, G.I., Pringle, J.E., 2002, MNRAS,
  337, 706

\bibitem{} Moeckel, N., Bally, J., 2006, ApJ, 653, 437

\bibitem{} Monnier, J.D., Berger, J.-P., Millan-Gabet, R., Traub,
  W.A., Schloerb, F.P., Pedretti, E., Benisty, M., Carleton, N.P.,
  Haguenhauer, P., Kern, P., Labeye, P., Lacasse, M.G., Malbet, F.,
  Perraut, K., Pearlman, M., Zhao, M., 2006, ApJ, 647, 444

\bibitem{} Muzerolle, J., Calvet, N., Hartmann, L., D'Alessio, P.,
  2003, ApJ, 597, L149

\bibitem{} Natta, A., Prusti, T., Neri, R., Wooden, D., Grinin, V.P.,
  Mannings, V., 2001, A\&A, 371, 186

\bibitem{} Ogilvie, G.I., 2006, MNRAS, 365, 977

\bibitem{} Pfalzner, S., Vogel, P., Scharw\"achter, J.,Olczak, C.,
  2005, A\&A, 437, 967

\bibitem{} Papaloizou, J.C.B., Lin, D.N.C., 1995, ApJ, 438, 841

\bibitem{} Pringle, J.E., 1981, ARA\&A, 19, 137

\bibitem{} Pringle, J.E., 1996, MNRAS, 281, 357

\bibitem{} Pringle, J.E., 1997, MNRAS, 292, 136 

\bibitem{} Pringle, J.E., 1999, in Astrophysical Discs,
  ed. J.A. Sellwood \& J. Goodman, Astr. Soc. Pac. Conf Ser. 160, 53

\bibitem{} Quillen, A.C., 2006, ApJ, 640, 1078

\bibitem{} Shakura, N.I., Sunyaev, R.A., 1973, A\&A, 24, 337

\bibitem{} Terquem, C.E.J.M.L.J., 2008, ApJ, 689, 532

\bibitem{} Terquem, C., Bertout, C., 1993, A\&A, 274, 291

\bibitem{} Terquem, C., Bertout, C., 1996, MNRAS, 279, 415

\bibitem{} Watkins, S.J., Bhattal, A.S., Biffin, H.M.J., Francis, N.,
  Whitworth, A.P., 1998a, MNRAS, 300, 1205

\bibitem{} Watkins, S.J., Bhattal, A.S., Biffin, H.M.J., Francis, N.,
  Whitworth, A.P., 1998b, MNRAS, 300, 1214

\end{thebibliography}
\end{document}